\documentclass[prb,twocolumn,showpacs]{revtex4}

\usepackage{graphicx}

\begin{document}

\title{Theory of proximity effect in 
superconductor/ferromagnet heterostructures}

\author{A.\ Bagrets}
\affiliation{
Laboratoire Louis N\'eel, CNRS, B.P.\ 166, 38042 Grenoble cedex 9, France}

\author{C.\ Lacroix}
\affiliation{
Laboratoire Louis N\'eel, CNRS, B.P.\ 166, 38042 Grenoble cedex 9, France}

\author{A.\ Vedyayev}
\affiliation{
CEA/Grenoble, D\'epartement de Recherche Fondamentale sur la
Mati$\grave e$re Condens\'ee,
SP2M/NM, 38054 Grenoble, France }
\affiliation{
Department of Physics, M.~V.~Lomonosov Moscow State University,
119899 Moscow, Russia }

\date{\today}

\begin{abstract}

We present a microscopic theory of proximity effect in the 
ferromagnet\-/superconductor\-/ferromagnet (F/S/F) nanostructures 
where S is $s$-wave low-$T_c$ superconductor and F's are 
layers of 3$d$ transition ferromagnetic metal.
Our approach is based on the direct analytical solution of 
Gor'kov equations for the normal and anomalous Green's functions
together with a self-consistent evaluation of the superconducting
order parameter. We take into account the elastic spin-conserving
scattering of the electrons assuming $s$-wave scattering in the S layer 
and $s$-$d$ scattering in the 
F layers. In accordance with 
the previous quasiclassical theories, we found that 
due to exchange field in the ferromagnet the anomalous Green's 
function $F(z)$ exhibits the damping oscillations in the F-layer
as a function of distance $z$ from the S/F interface. In the given 
model a half of period of oscillations is determined by the 
length $\xi_m^0 = \pi v_F/\varepsilon_{\rm ex}$, where $v_F$ is the
Fermi velocity and $\varepsilon_{\rm ex}$ is the exchange field, 
while damping is governed by the length 
$l_0 = (1/l_{\uparrow} + 1/l_{\downarrow})^{-1}$ with $l_{\uparrow}$ 
and $l_{\downarrow}$ being spin-dependent mean free paths in the
ferromagnet. The superconducting transition temperature $T_c(d_F)$
of the F/S/F trilayer shows the damping oscillations
as a function of the F-layer thickness $d_F$ with period
$\xi_F = \pi/\sqrt{m\varepsilon_{\rm ex}}$, where
$m$ is the effective electron mass. The oscillations of $T_c(d_F)$ are 
a consequence 
of the oscillatory behavior of the superconducting order
parameter at the S/F interface {\it vs}\ thickness $d_F$, 
that in turn is caused by the oscillations of $F(z)$ in the F-region.
We show that strong spin-conserving scattering either in the superconductor 
or in the ferromagnet significantly suppresses these oscillations.
The calculated $T_c(d_F)$ dependences are compared with existing 
experimental data for Fe/Nb/Fe trilayers and Nb/Co multilayers.  

\end{abstract}



\maketitle

\section{Introduction}

The artificially fabricated layered nanostuctures with 
alternating superconducting (S) and ferromagnetic (F) 
layers provide a possibility to study 
the physical phenomena arising due to proximity of
two materials (S and F) with two antagonistic types of 
long range ordering. One of such interesting effects
is the existence of so-called $\pi$-phase superconducting state 
in which the order parameter in adjacent S-layers has opposite sings.
The $\pi$-junctions were originally predicted 
to be possible due to spin-flip processes in magnetic 
layered structures containing paramagnetic impurities in the barrier 
between S layers \cite{Bulaevskii}. Later on, Buzdin {\it et al.}
\cite{Buzdin_JETPLett,Buzdin_JETP} and Radovi\'c {\it et al.} \cite{Buzdin_PRB} 
showed that, due to oscillatory behavior of the Cooper pair 
wave function in the ferromagnet, $\pi$-coupling can be realized also 
for S/F multilayers. The $\pi$-coupling leads to a 
nonmonotonic oscillatory dependence of the superconducting transition 
temperature $T_c$ as a function of ferromagnetic layer 
thickness~$d_F$\cite{Buzdin_JETPLett,Buzdin_JETP,Buzdin_PRB}.
The effect occurs because of periodically switching of the ground state 
between $0$- and $\pi$-phase, so that the system chooses the state with 
higher transition temperature $T_c$. 

   These theoretical predictions
stimulated a considerable interest to proximity effect in S/F 
structures also from experimental point of view. First the oscillatory 
behavior of $T_c(d_F)$ was observed by Wong {\it et al.} \cite{Wong} in 
V/Fe multilayers and later on these results were well explained by  
theoretical calculations of Radovi\'c {\it et al.} \cite{Buzdin_PRB} 
However, in subsequent experiments with V/Fe multilayers \cite{Koorevaar}, 
the oscillatory $T_c(d_F)$ dependence was not observed. 
The following experiments\cite{Jiang,Muhge,Strunk,Muhge1,Verbanck,Obi,Aarts} 
revealed 
the different and even controversial behavior of $T_c(d_F)$ for different 
structures. The nonmonotonic oscillation-like behavior of $T_c(d_F)$ was 
reported by Jiang {\it et al.} \cite{Jiang} for Nb/Gd multilayers and
Nb/Gd/Nb trilayers, by M\"uhge {\it et al.} \cite{Muhge} for Fe/Nb/Fe 
trilayers and recently by Obi {\it et al.} \cite{Obi} for Nb/Co and V/Co 
multilayers. However, negative results were reported for Nb/Gd/Nb 
trilayers by Strunk {\it et al.} \cite{Strunk}, 
for V/V$_{1-x}$Fe$_x$ multilayers by Aarts {\it et al.} \cite{Aarts},
for Fe/Nb bilayers by M\"uhge {\it et al.} \cite{Muhge1} and
Nb/Fe multilayers by Verbanck {\it et al.} \cite{Verbanck}.
For interpretation of experimental results, along with 
mechanisms of $\pi$-coupling and suppression of $T_c$
due to strong exchange field in the ferromagnet, 
another mechanisms were suggested such as  complex behavior
of the "magnetically dead" interfacial S/F layer 
(see details in Ref.~\onlinecite{Muhge}), 
the effects of a finite interface transparency \cite{Aarts}, 
and spin-orbit scattering \cite{SO}.

The original theory of proximity effect proposed by Buzdin 
{\it et al.} \cite{Buzdin_JETP,Buzdin_PRB} is based
on the quasiclassical Usadel equations \cite{Usadel}
applied for S/F structures. In this case 
the Usadel equations must be supplemented by boundary conditions 
for the quasiclassical Green's functions
at the S/F interface. This essential point was recently discussed 
in Ref.~\onlinecite{Khusainov}.
On the other hand,
 the boundary conditions for microscopic Green's functions
can be written obviously
for ideal S/F interfaces if one uses Gor'kov equations \cite{Gorkov}. 
These equations, however, are more complex to resolve than the quasiclassical 
ones. In the given paper, we present a theoretical investigation 
of $T_c(d_F)$ behavior for F/S/F trilayer structures based  
on Gor'kov equations. We consider that F layers are 
3$d$ transition metals and assume that the main mechanism
of spin-conserving electron scattering in F layers is $s$-$d$ scattering, 
while S layer is $s$-wave superconductor with $s$-$s$ scattering. 
We find the characteristic lengths determining
the periods of oscillations and damping of critical 
temperature $T_c$ and Cooper pair wave function,
and show that in the given 
model
these lengths differ from length scales predicted 
by quasiclassical theories \cite{Buzdin_JETP,Buzdin_PRB,Demler,Khusainov}. 
We show that strong spin-conserving scattering either in the superconductor 
or in the ferromagnet significantly suppresses the oscillations of $T_c$.
We compare our results with existing data
on $T_c(d_F)$ for Fe/Nb/Fe trilayers \cite{Muhge}
and V/Co multilayers \cite{Obi}, where F's are 3$d$ ferromagnets,
and find reasonable agreement with theory and experiment.

\section{Gor'kov equations and Green's functions}

We  consider a trilayer structure F$_1$/S/F$_3$ where S is 
low-$T_c$ superconductor and F's are 3$d$-metal ferromagnetic layers.
The thicknesses $d_S$ and $d_F$ 
of the S- and F-layers are supposed to be much smaller than the in-plane
dimension of the structure, so that the system can be considered 
as homogeneous in the $xy$-plane (parallel to the interfaces). 
We denote the axis perpendicular to the $xy$-plane as $z$-axis. 
Let $z = \pm a$ be the positions of the outer boundaries 
of the F-layers and $z = \pm d$ be the positions of S/F interfaces, 
then $d_S = 2d$ and $d_F = a-d$. 
We adopt that S is a simple $s$-wave superconductor 
with $s$-$s$ mechanism of electron scattering.  
According to Ref.~\onlinecite{Weber}, for superconducting Nb 
which is usually used in preparing the S/F heterostructures,  
\mbox{$s$-wave} scattering is indeed prevailing. 
Concerning the ferromagnetic layers, 
we adopt the 
simplified model\cite{Ehrenreich} considering 
that two types of electrons form the total band 
structure of 3$d$ transition metals: almost free-like spin-up
and spin-down electrons from $sp$ bands (these electrons are 
referred as $s$ electrons) and localized $d$ electrons 
from narrow strongly exchange split bands. The main mechanism of 
spin-conserving electron scattering in 3$d$ ferromagnetic metals is  
$s$-$d$ scattering \cite{Brouers} because of a dominant contribution 
of $d$ density of states (DOS) to the total DOS at the Fermi 
energy $\varepsilon_F$. The mean free path of the conduction $s$ electrons 
depends on the spin due to $s$-$d$ scattering and the different 
$d$ density of states at $\varepsilon_F$ for majority and minority spin 
bands. In the present work we consider only
the scattering on nonmagnetic impurities.

As a starting point, we take the system 
of Gor'kov equations \cite{Gorkov} 
for the normal and anomalous Green's functions 
$G^{ss}_{\uparrow\uparrow}(x_1, x_2) = - 
\bigl \langle T_{\tau} \psi_{\uparrow}(x_1) 
\psi_{\uparrow}^{\dagger}(x_2) \bigr \rangle$
and 
$F^{ss}_{\downarrow\uparrow}(x_1, x_2) =  
\bigl \langle T_{\tau} \psi_{\downarrow}^{\dagger}(x_1) 
\psi_{\uparrow}^{\dagger}(x_2) \bigr \rangle$,
where $x = (\tau, \mathbf{r})$ is a four-component vector
and the creation and annihilation field operators are associated 
with $s$ electrons. By carrying out the Fourier transformation 
in the $xy$-plane and over the imaginary time $\tau$, 
we get the following system for the Green's functions: 

i) for the F-layers:
\begin{eqnarray}
\Biggl [ i\omega + \frac{1}{2m}\left( 
\frac{\partial ^2}{\partial z^2} - \kappa^2 \right) + 
\varepsilon_F + h(z) 
- x_0 \gamma^2_{sd} G^{dd}_{\uparrow\uparrow}(z,z) \Biggr ] 
\nonumber  \\
\times\ G^{ss}_{\uparrow\uparrow}(z,z') + 
\Delta(z) F^{ss}_{\downarrow\uparrow}(z,z') = \delta(z - z'),
\nonumber \\
\Delta^*(z) G^{ss}_{\uparrow\uparrow}(z,z') + 
\Biggl [ i\omega - \frac{1}{2m}\left( 
\frac{\partial ^2}{\partial z^2} - \kappa^2 \right)
\qquad \qquad \ \ 
\label{Gor_F} \\
-\ \varepsilon_F + h(z) - 
x_0 \gamma^2_{sd} G^{dd}_{\downarrow\downarrow}(z,z) \Biggr ] 
F^{ss}_{\downarrow\uparrow}(z,z') = 0,
\nonumber
\end{eqnarray}

\nonumber
ii) for the S-layer:
\begin{eqnarray}
\Biggl [ i\omega' + \frac{1}{2m}\left( 
\frac{\partial ^2}{\partial z^2} - \kappa^2 \right) 
+ \varepsilon_F \Biggr ] 
G^{ss}_{\uparrow\uparrow}(z,z') \qquad \qquad \quad 
\nonumber \\
+\ \Delta_{\omega}(z) F^{ss}_{\downarrow\uparrow}(z,z') = \delta(z - z'),
\nonumber  \\
\Delta^*_{\omega}(z) G^{ss}_{\uparrow\uparrow}(z,z') +
\hspace{4.8cm} 
\label{Gor_S} \\
\Biggl [ i\omega' - \frac{1}{2m}\left( 
\frac{\partial ^2}{\partial z^2} - \kappa^2 \right) 
- \varepsilon_F \Biggr ] 
F^{ss}_{\downarrow\uparrow}(z,z') = 0 
\nonumber
\end{eqnarray}
with
\begin{eqnarray}
\omega'& = & \omega + i c u_0^2 G^{ss}_{\uparrow\uparrow}(z,z), 
\nonumber \\
\Delta_{\omega}(z)& = & \Delta(z)
 + cu_0^2 F^{*ss}_{\downarrow\uparrow}(z,z).
\label{sc_S}
\end{eqnarray}

In Eqs.\ (\ref{Gor_F})--(\ref{Gor_S}) $\kappa$ is the in-plane
momentum, parallel to the S/F interface, $m$ is the effective
electron mass which is assumed to be the same for both metals,
$h(z)$ is exchange field in the ferromagnet, 
$\omega = \pi T (2n + 1)$ are 
Matsubara frequencies (the units are $\hbar = 1 = k_B$). 
The scattering processes are introduced in the Born approximation.
The parameters $u_0$ and $\gamma_{sd}$ are the strengths of 
impurity potentials, $c$ and $x_0$ are impurity concentrations
in the S- and F-layers. We assume that a BCS coupling constant 
is zero for the ferromagnet, therefore $\Delta(z) = 0$ 
in the F-layers. We also neglect by the 
possible deviation of $\Delta(z)$ from zero 
in the F-region due to scattering, since this correction is of the 
order of $\gamma_{sd}^4$ which is small.

  The superconductor order parameter has to be found self-consistently, 
\begin{equation}
\Delta(z) = \lambda\ T \sum_{\omega} \int_{0}^{k_F} 
\frac{\kappa d\kappa}{2\pi}\ F^*(\omega, \kappa, z = z'),
\label{self-cons}
\end{equation}
where summation over $\omega$ goes up to Debye frequency $\omega_D$, 
$\lambda > 0$ is the BCS coupling constant in a superconductor,
and $F = F_{\downarrow\uparrow}^{ss}$. 
The critical temperature $T_c$ is defined as the first zero of  
equation $\Delta(z) = 0$ when $T$ decreases from high temperatures.

Below in this section and in Sec.~III
we present a scheme to evaluate the Green's functions
considering as the first step the non-self-consistent
solution of Eqs.~(\ref{Gor_F},\ref{Gor_S}) 
where $\Delta(z) = \Delta$ is a real number which does
not depend on $z$. Sec.~IV is devoted to the self-consistent 
evaluation of 
$\Delta(z)$. We will assume, that
the mutual orientation of magnetizations in the F layers
is antiparallel (AP), therefore $h(z) = h > 0$ in the F$_1$-layer, 
and $h(z) = -h$ in the F$_3$-layer.
 The advantage of 
the AP configuration is that in this case 
the self-consistency can be achieved for real 
values of $\Delta(z)$ in the S region. 
The study of the influence
of the mutual orientation of magnetizations on $T_c$
(Refs.~\onlinecite{Vedyayev,Tagirov,Buzdin_new}) in the framework of
the given model requires to consider $\Delta(z)$ as a complex
valued function. These question is beyond the present
study and will be discussed in the forthcoming publication.
However, as can be seen further, 
the general conclusions of the given paper
are not sensitive to the particular configuration 
of the magnetizations.
At the first step we also suppose that there is no
scattering in the S layer. The scattering processes
in the S layer (Eq.~\ref{sc_S}) are taken into account
at the last step of the evaluation of the critical temperature
(Sec.~V).

By introducing the Green's functions
$\widetilde G^{ss}_{\downarrow\downarrow}(x_1, x_2) = - 
\bigl \langle T_{\tau} \psi_{\downarrow}^{\dagger}(x_1) 
\psi_{\downarrow}(x_2) \bigr \rangle$
and 
$\widetilde F^{ss}_{\uparrow\downarrow}(x_1, x_2) =  
\bigl \langle T_{\tau} \psi_{\uparrow}(x_1) 
\psi_{\downarrow}(x_2) \bigr \rangle$
the system of Gor'kov equations can be
written in the matrix form \cite{Svidz}
\begin{equation}
\Bigl[ 
i\omega \hat I - \hat {\cal A} \Bigr]
\left( 
\begin{array}{cc}
G & \widetilde F \\
F & \widetilde G
\end{array}
\right)  =  \hat I \delta(z - z'), 
\label{Gor_matrix}
\end{equation}
where $\hat I$ is the unit matrix, and
$\hat{\cal A}$ is the $(2\times 2)$-matrix differential 
operator, the components of which can be found 
by comparing the Eqs.\ (\ref{Gor_F})--(\ref{Gor_S}) and 
Eq.\ (\ref{Gor_matrix}).
  
  In order to find the matrix Green's function,
consider the Schr\"odinger's equation with the 
Hamiltonian $\hat{\cal A}$:
\begin{equation}
\Bigl[ i\omega \hat I - \hat{\cal A} \Bigr] \psi(z) = 0.   
\label{Schrod}
\end{equation}
This equation has four linear independent solutions, 
\[
\varphi_{\mu}(z) = \left(
\begin{array}{c} 
\varphi_{\mu}^+(z) \\ \varphi_{\mu}^-(z) 
\end{array}
\right) \quad (\mu = \uparrow, \downarrow),
\]
and
\[
\psi_{\rho}(z) = \left(
\begin{array}{c} 
\psi_{\rho}^+(z) \\ \psi_{\rho}^-(z) 
\end{array}
\right) \quad (\rho = \uparrow, \downarrow).
\]
We require that $\psi_{\mu}(z)$ and $\psi_{\rho}(z)$
obey zero boundary conditions at the points
$z = \pm a$, and choose these independent solutions 
in such a way that two functions $\varphi_{\uparrow}(z)$ and 
$\psi_{\uparrow}(z)$ describe spin-up electrons
in the ferromagnetic layers, and functions $\varphi_{\downarrow}(z)$ 
and $\psi_{\downarrow}(z)$ describe spin-down holes in the F-layers. 
Namely, in the layer F$_1$ $(-a<z<-d)$ the solutions $\varphi_{\mu}(z)$ 
have the form
\begin{eqnarray}
\varphi_{\uparrow}(z) & = & \left(1 \atop 0 \right) 
\sin \left[ p_1^{\uparrow} (z + a)  \right],
\label{sol_F_1} \\
\varphi_{\downarrow}(z) & = & \left(0 \atop 1 \right) 
\sin \left[ p_1^{\downarrow} (z + a)  \right],
\nonumber
\end{eqnarray}
and in the layer F$_3$ $(d<z<a)$ the solutions $\psi_{\rho}(z)$ are 
\begin{eqnarray}
\psi_{\uparrow}(z) & = & \left(1 \atop 0 \right) 
\sin \left[ p_3^{\uparrow} (a - z)  \right],
\label{sol_F_2} \\
\psi_{\downarrow}(z) & = &\left(0 \atop 1 \right) 
\sin \left[ p_3^{\downarrow} (a - z)  \right].
\nonumber
\end{eqnarray}
Here $p_1^{\uparrow(\downarrow)}$
is an electron (hole) momentum in the layer~F$_1$,
\begin{equation}
p_1^{\uparrow(\downarrow)} = 
\sqrt{2m\left(\varepsilon_F - \frac{\kappa^2}{2m} 
\pm h \pm \frac i2\ \tau_{\uparrow(\downarrow)}^{-1} \pm i\omega 
\right) }
,
\label{p1}
\end{equation}
and $p_3^{\uparrow(\downarrow)}$ are momenta in the layer~F$_3$, 
\begin{equation}
p_3^{\uparrow(\downarrow)} = 
\sqrt{2m\left(\varepsilon_F - \frac{\kappa^2}{2m} 
\mp h \pm \frac i2\ \tau_{\downarrow(\uparrow)}^{-1} 
\pm i\omega \right) }.
\label{p3}
\end{equation}
The inverse life-times of 
quasiparticles are 
given by
$\tau_{\uparrow(\downarrow)}^{-1} = 
- 2 x_0 \gamma_{sd}^2 \mathrm{Im} 
G^{dd}_{\uparrow\uparrow(\downarrow\downarrow)} = 
k_F^{\uparrow(\downarrow)}/(m l_{\uparrow(\downarrow)})$, 
here 
$k_F^{\uparrow(\downarrow)} = \sqrt{2m(\varepsilon_F \pm h )}$ 
being Fermi momenta in the ferromagnet and 
$l_{\uparrow(\downarrow)}$ being mean free paths
which are considered as parameters. 

  In the S-region $(-d < z < d)$ the solutions of 
Eq.~(\ref{Schrod}) are 
\begin{eqnarray}
\varphi_{\mu}(z) & = 
\displaystyle
A^{\mu}_{+} \left(1 \atop \alpha \right) e^{ik_{+}(z+d)} + 
A^{\mu}_{-} \left(1 \atop \alpha \right) e^{-ik_{+}(z+d)} 
\nonumber \\
& \displaystyle + \
B^{\mu}_{+} \left( \alpha \atop 1 \right) e^{ik_{-}(z+d)} + 
B^{\mu}_{-} \left(\alpha \atop 1 \right) e^{-ik_{-}(z+d)},
\label{sol_S} \\
\psi_{\rho}(z) & = 
\displaystyle
C^{\rho}_{+} \left(1 \atop \alpha \right) e^{ik_{+}(z - d)} + 
C^{\rho}_{-} \left(1 \atop \alpha \right) e^{-ik_{+}(z - d)} 
\nonumber \\
& \displaystyle + \
D^{\rho}_{+} \left( \alpha \atop 1 \right) e^{ik_{-}(z - d)} + 
D^{\rho}_{-} \left(\alpha \atop 1 \right) e^{-ik_{-}(z - d)},
\nonumber
\end{eqnarray}
where the wave vectors $k_{\pm}$ are defined as
\[
k_{\pm} = \sqrt{2m\left( 
\varepsilon_F - \frac{\kappa^2}{2m} \pm 
i\sqrt{\omega^2 + \Delta^2} \right)}, 
\]
and 
\[
\alpha = \frac{i}{\Delta}
\left[ \sqrt{\omega^2 + \Delta^2} - \omega \right].
\]
We neglect the interfacial roughness, 
thus the coefficients 
$A_{\pm}^{\mu}$,  $B_{\pm}^{\mu}$,  
$C_{\pm}^{\rho}$,  $D_{\pm}^{\rho}$ have to be found
from the conditions of continuity of the functions 
$\varphi_{\mu}(z)$ and $\psi_{\rho}(z)$ and 
their derivatives at the points $z = \pm d$, 
that can be done easily by solving the system
of algebraic linear equations.

 To evaluate the matrix Green's function, let
us introduce the matrices 
\[
\Phi(z) =
\left(
\begin{array}{cc}
\varphi_{\uparrow}^{+}(z) & \varphi_{\downarrow}^{+}(z) \\
\varphi_{\uparrow}^{-}(z) & \varphi_{\downarrow}^{-}(z)
\end{array}
\right),
\]
\[
\Psi(z') =
\left(
\begin{array}{cc}
\psi_{\uparrow}^{+}(z') & \psi_{\downarrow}^{+}(z') \\
\psi_{\uparrow}^{-}(z') & \psi_{\downarrow}^{-}(z')
\end{array}
\right),
\]
and let $J$ be the matrix of "currents",
\[
J = \left(
\begin{array}{cc}
j_{\uparrow\uparrow} & j_{\uparrow\downarrow} \\
j_{\downarrow\uparrow} & j_{\downarrow\downarrow}
\end{array}
\right),
\]
with components
\begin{equation}
j_{\mu\rho} = 
\varphi_{\mu}^{+}(z) \stackrel{\leftrightarrow}{\nabla}_z 
\psi_{\rho}^{+}(z) -
\varphi_{\mu}^{-}(z) \stackrel{\leftrightarrow}{\nabla}_z 
\psi_{\rho}^{-}(z),
\label{currents}
\end{equation}
here 
$\mu, \rho = \uparrow, \downarrow$, and
$\stackrel{\leftrightarrow}{\nabla}_z = 
(\stackrel{\rightarrow}{\nabla}_z  - 
\stackrel{\leftarrow}{\nabla}_z) 
$ is the antisymmetric gradient operator.
The matrix $J$ is the Wronskian of the system 
(\ref{Schrod}) which does not depend on $z$, 
i.e., $\partial J(z)/\partial z = 0$.
Finally, the matrix Green's function introduced in 
(\ref{Gor_matrix}) is given by
\begin{equation}
\hat G(z,z') = 2m\
\Phi(z) [J^{-1}]^T \Psi(z').
\label{matr_Green}
\end{equation}
Here $T$ denotes the transposition operation.
The obtained expression allows
to evaluate the normal and anomalous 
Green's functions in both layers (S and F).

\section{Anomalous Green's function}

\subsection{S-layer}

Consider first the anomalous Green's 
function (Cooper pair wave function) 
$F(\omega, \kappa, z=z')$
in the S-region.  Denote $\theta_{\pm} = 2ik_{\pm}d$, and  
$\theta_{\pm} = \theta \pm i\delta$, i.e., $\theta$ and 
$\delta$ are real and imaginary parts of phases $\theta_{\pm}$.
Using solutions (\ref{sol_S}) of Eq.~(\ref{Schrod}) in the superconductor, 
we get the exact expressions for currents
\[
j_{\mu\rho} = (1 - \alpha^2) j_{\mu\rho}^0,
\]
where 
\begin{eqnarray}
j_{\mu\rho}^0 & = & 2ik_{+} \left[ 
A_{-}^{\mu} C_{+}^{\rho} e^{-i\theta_{+} } - 
A_{+}^{\mu} C_{-}^{\rho} e^{i\theta_{+} } \right] 
\label{j_sc} \\
& - & 2ik_{-} \left[ 
B_{-}^{\mu} D_{+}^{\rho} e^{-i\theta_{-} } - 
B_{+}^{\mu} D_{-}^{\rho} e^{i\theta_{-} }
\right].  
\nonumber 
\end{eqnarray}
Since the currents $j_{\mu\rho}$ do not depend on $z$, 
the same expressions can be obtained using the solutions 
of Eq.~(\ref{Schrod}) in the ferromagnetic layers. 

 It is convenient to introduce an energy variable 
$\xi = \varepsilon_F - \kappa^2/2m$. The typical dependence
of $F(\xi)$ on $\xi$ under given arguments $\omega$ and $\Delta$
at the point $z = z' = 0$ is shown in Fig.~1.
Function $F(\xi)$ exhibits the quantum oscillations which 
are the result of exponentials $e^{\pm i\theta_{\pm}}$ 
in Eq.~(\ref{j_sc}) with
rapidly varying phases. Since the superconducting 
order parameter is determined by the integral of $F(\xi)$ over
$\xi$, one can average $F(\xi)$ over the oscillations.  

  Denote $a_{\mu\rho}$ ($\mu, \rho = \uparrow, \downarrow$)
the components of the matrix 
\[
[J^{-1}]^T = 
\left(
\begin{array}{cc}
a_{\uparrow\uparrow} & a_{\uparrow\downarrow} \\
a_{\downarrow\uparrow} & a_{\downarrow\downarrow}
\end{array}
\right).
\]
For $a_{\mu\rho}$ we get
\[
a_{\mu\rho} = \frac{\mathrm{sign}(\mu\rho)}{(1-\alpha^2)\mathrm{Den}}
j_{-\mu,-\rho}^0  
= \frac{1}{\mathrm{Den}} 
\Bigl[ a_{\mu\rho}^{-}e^{-i\theta} + 
a_{\mu\rho}^{+}e^{i\theta} \Bigr], 
\]
where
\begin{eqnarray}
a_{\mu\rho}^{-} & = & 
\frac{\mathrm{sign}(\mu\rho)}{(1-\alpha^2)}
\Bigl[ 
2ik_{+} A_{-}^{-\mu} C_{+}^{-\rho} e^{\delta} 
\nonumber \\
& - & 
2ik_{-} B_{-}^{-\mu} D_{+}^{-\rho} e^{-\delta}
\Bigr], 
\nonumber \\
a_{\mu\rho}^{+} & = &
- \frac{\mathrm{sign}(\mu\rho)}{(1-\alpha^2)}
\Bigl[ 
2ik_{+} A_{+}^{-\mu} C_{-}^{-\rho} e^{-\delta} 
\label{a_coeff} \\
& - &  2ik_{-} B_{+}^{-\mu} D_{-}^{-\rho} e^{\delta}
\Bigr],   
\nonumber
\end{eqnarray}
here $\theta \pm i\delta = \theta_{\pm} = 2ik_{\pm}d$, and  
$\mathrm{Den} = \mathrm{det} J /(1-\alpha^2)^2$ is the
determinant of the matrix of currents:
\[
\mathrm{Den} = - D_0 + 
\Gamma_{+} e^{2i\theta} + \Gamma_{-} e^{-2i\theta}.
\]
The expressions for $D_0$ and $\Gamma_{\pm}$ are given 
in Appendix~A.

 By carrying out the Fourier transformation 
of $a_{\mu\rho}$, 
we can write the first terms of the expansion:
\begin{equation}
\langle a_{\mu\rho} \rangle = 
b_{\mu\rho}^{+} e^{i\theta} + 
b_{\mu\rho}^{-} e^{-i\theta} + \dots ,
\label{Four_a}
\end{equation}
where $b_{\mu\rho}^{\pm}$ are defined by the following integrals
\[
b_{\mu\rho}^{\pm} = 
\frac{1}{2\pi} \int_{-\pi}^{+\pi} d\phi\
\frac{a_{\mu\rho}^{\mp} e^{-i\phi} + a_{\mu\rho}^{\pm}}
{\Gamma_{+} e^{i\phi} + \Gamma_{-} e^{-i\phi} - D_0}.
\]

\begin{figure}[t]
\begin{center}
\includegraphics[scale = 0.35]{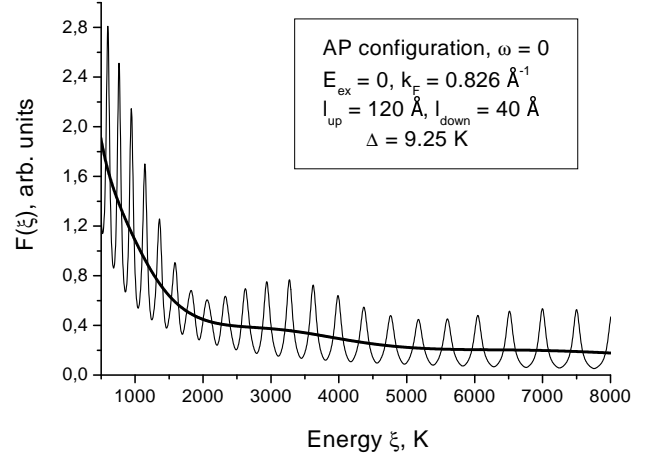}
\caption{The typical dependence of the anomalous Green's function
$F(\omega, \xi, z)$ on the energy $\xi = \kappa^2/2m -
\varepsilon_F$ under given $\Delta = 9.25$~K, $\omega = 0$,
$z = 0$ and $\varepsilon_{\rm ex} = 0$,
mean free paths in the F-layer are $l_{\uparrow} = 120$~\AA\
and  $l_{\downarrow} = 40$~\AA, Fermi momentum
$k_F = 0.826$~\AA$^{-1}$. $F(\xi)$ exhibits quantum oscillations,
the smooth line shows function $F(\xi)$ averaged over oscillations,
i.e.\ $F_0(\xi)$ (see the text for the details). }
\end{center}
\end{figure}

 Using Eq.~(\ref{matr_Green}), we get expression
for $F(\omega, \xi, z = z')$:
\[
F(\omega, \xi, z = z') \ =  \
2m \sum_{\mu,\rho = \uparrow\downarrow} 
\varphi^{-}_{\mu}(z)\langle a_{\mu\rho}\rangle
\psi^{+}_{\rho}(z)
\]
where $\varphi^{-}_{\mu}(z)$ and $\psi^{+}_{\rho}(z)$ are 
the components of solutions 
$\varphi_{\mu}(z)$ and $\psi_{\rho}(z)$ in the S-layer.
Denote
\[
\theta_1 \pm i\delta_1 = k_{\pm}(d + z), \quad
\theta_2 \pm i\delta_2 = k_{\pm}(d - z), 
\]
then $\theta_1 + \theta_2 = \theta$, $\delta_1 + \delta_2 = \delta$
and 
\begin{eqnarray}
\varphi_{\mu}^{-}(z) & = &
\Lambda_{+}^{\mu}e^{i\theta_1} + 
\Lambda_{-}^{\mu}e^{-i\theta_1}, 
\nonumber \\ 
\psi_{\rho}^{+}(z) & = & 
\Sigma_{+}^{\rho}e^{i\theta_2} + 
\Sigma_{-}^{\rho}e^{-i\theta_2},
\nonumber
\end{eqnarray}
where
\begin{eqnarray}
\Lambda_{+}^{\mu} = 
\alpha A_{+}^{\mu} e^{-\delta_1} + B_{+}^{\mu} e^{\delta_1}, & \ &
\Lambda_{-}^{\mu} = 
\alpha A_{-}^{\mu} e^{\delta_1} + B_{-}^{\mu} e^{-\delta_1}, 
\nonumber \\
\Sigma_{+}^{\rho} = 
C_{-}^{\rho} e^{-\delta_2} + \alpha D_{-}^{\rho} e^{\delta_2}, & \ &
\Sigma_{-}^{\rho} = 
C_{+}^{\rho} e^{\delta_2} + \alpha D_{+}^{\rho} e^{-\delta_2}.
\nonumber
\end{eqnarray}

The higher order terms with $e^{\pm i n \theta}$, where $n \ge 2$,
can be dropped 
in the expansion (\ref{Four_a})
 because they are responsible for 
rapid oscillations of $F(\xi)$. Finally, we come to the 
following expression for the anomalous Green's function 
in the S-layer,
\[
F(\omega, \xi, z = z')  =  
F_0(\omega, \xi, z) + 
F_1(\omega, \xi, z),
\]
where
\begin{eqnarray}
F_0(\omega, \xi, z) &  = &  2m
\sum_{\mu\rho} \Bigl\{ 
\Lambda_{+}^{\mu} \Sigma_{+}^{\rho} b_{\mu\rho}^{-} +
\Lambda_{-}^{\mu} \Sigma_{-}^{\rho} b_{\mu\rho}^{+} \Bigr\},
\label{F0} \\
F_1(\omega, \xi, z) &  = &  2m
\sum_{\mu\rho} \Bigr\{
\Lambda_{+}^{\mu} \Sigma_{-}^{\rho} \Bigl[ 
b_{\mu\rho}^{+} e^{2i\theta_1} + b_{\mu\rho}^{-} e^{-2i\theta_2} 
\Bigr]
\nonumber \\
& + &
\Lambda_{-}^{\mu} \Sigma_{+}^{\rho} \Bigl[ 
b_{\mu\rho}^{+} e^{2i\theta_2} + b_{\mu\rho}^{-} e^{-2i\theta_1} 
\Bigr] \Bigr\}.
\nonumber
\end{eqnarray}

The contribution $F_1$ to the function $F$ is essential 
only in the vicinity of S/F interfaces, $z = \pm d$, as far as 
$\theta_1 \sim (z + d)$ and $\theta_2 \sim (d - z)$. At the point $z = z' = 0$ 
(the middle of the S-layer) the anomalous Green's function is determined by the
function $F_0(\omega, \xi, z)$ which is shown by the 
thick smooth line in Fig.~1. The obtained result 
is used below in Sec.~IV 
where we discuss the self-consistent 
evaluation of the order parameter.

\subsection{F-layer}

Due to proximity effect, the 
correlations between electrons are induced in the ferromagnet 
close to the supercoducting layer. Instead of simple 
decay, as it would be for the 
super\-conductor/\-normal-metal interface, 
in the case of ferromagnetic layer the Cooper pair 
wave function exhibits the damping oscillatory behavior 
in the ferromagnet with increasing a distance from the S/F 
interface \cite{Buzdin_JETP,Buzdin_PRB,Demler}. 
The reason is that exchange splitting 
of bands in the F-region changes 
the pairing conditions for electrons, therefore the Cooper pairs
are formed from quasiparticles with equal energies but
with different in modulus momenta $p_{\uparrow}$ and $-p_{\downarrow}$.
Due to the non-zero center of mass momentum $\Delta p$, 
the Cooper pair wave function obtains the spatially 
dependent phase in the ferromagnetic layer.  
In the "clean" limit (no scattering in the ferromagnet) 
one can find \cite{Demler} that the Cooper pair wave function 
oscillates 
with the distance $z$ into the F-layer as 
$\sim \sin (z/\xi_F^0)/ (z/\xi_F^0)$ where 
$\xi_F^0 = v_F/\varepsilon_{\rm ex}$.

This result holds also in the case of "dirty" ferromagnet. 
The microscopic theory of S/F multilayers based on 
the quasiclassical Usadel equations \cite{Buzdin_JETP,Buzdin_PRB} 
predicts that the anomalous Green's function behaves in the ferromagnet
as $\sim \exp\{-(1 + i)\sqrt{h/D_M} z \}$, where 
$D_M = v_F l / 3$ is the diffusion coefficient 
and $l$ is the electron mean free path in the F-layer.
Therefore, a length scale for oscillations and damping is 
the same and this scale is set by the length $\xi_M = \sqrt{2l\xi_F^0/3}$.
Below in this section it is shown that 
in the framework of our model  
the scales for oscillations and damping of the anomalous 
Green's function are determined by different lengths.  
 
  We can find the anomalous Green's function $F(\omega, \xi, z = z')$ 
in the F-region $(d < z < a)$ following the same approach that 
was used to evaluate the $F$-function in the superconductor. 
The solutions $\psi_{\uparrow(\downarrow)}(z)$ 
in the layer F$_3$ $(d < z < a)$ are given by 
Eq.~(\ref{sol_F_2}). For the solutions 
$\varphi_{\mu}(z)$ ($\mu = \uparrow, \downarrow$) we can
write   
\begin{eqnarray}
\varphi_{\mu}(z) & = 
\displaystyle
X^{\mu}_{+} \left(1 \atop 0 \right) e^{ip_3^{\uparrow}(a - z) } + 
X^{\mu}_{-} \left(1 \atop 0 \right) e^{-ip_3^{\uparrow}(a - z) } 
\nonumber \\
& \displaystyle + \
Y^{\mu}_{+} \left( 0 \atop 1 \right) e^{ip_3^{\downarrow}(a - z) } + 
Y^{\mu}_{-} \left( 0 \atop 1 \right) e^{-ip_3^{\downarrow}(a - z) }, 
\nonumber
\end{eqnarray}
where $X^{\mu}_{\pm}$ and $Y^{\mu}_{\pm}$ can be found
from conditions of continuity of the functions 
$\varphi_{\mu}(z)$ and their derivatives at $z=d$, 
assuming perfect S/F interface.

 The anomalous Green's function averaged over oscillations is
\begin{eqnarray}
F(\omega, \xi, z = z') & = & 2m \sum_{\mu} 
\varphi_{\mu}^{-}(z) \langle a_{\mu \uparrow} \rangle
\psi_{\uparrow}^{+}(z) 
\nonumber \\
& = &
2m \sum_{\mu} F_{\mu} \langle a_{\mu \uparrow} \rangle.
\label{F_ferro}
\end{eqnarray}
It turns out that function $\varphi_{\mu}^{-}(z)$ contains four terms
with multipliers $e^{\pm \theta_{+}}$ and 
$e^{\pm \theta_{-}}$. Denoting $\theta_{\pm} = \theta \pm i\delta$,
we can write $F_{\mu}$ in the form
\begin{equation}
F_{\mu} = \Phi_{\mu}^{+} e^{i\theta} + 
\Phi_{\mu}^{-} e^{-i\theta}, 
\label{F_index}
\end{equation}
where 
\begin{eqnarray}
\Phi_{\mu}^{+} & = & 
\Bigl[ \Theta_{+}^{\mu} \cos (p_3^{\downarrow}z_1) + 
\Xi_{+}^{\mu} \sin (p_3^{\downarrow}z_1 ) \Bigr] 
\nonumber \\
& \times &
\sin [ p_3^{\uparrow}(d_F - z_1) ],
\label{Phi} \\
\Phi_{\mu}^{-} & = &
\Bigl[ \Theta_{-}^{\mu} \cos (p_3^{\downarrow}z_1) + 
\Xi_{-}^{\mu} \sin (p_3^{\downarrow}z_1 ) \Bigr]
\nonumber \\
& \times &
\sin [ p_3^{\uparrow}(d_F - z_1) ],
\nonumber
\end{eqnarray}
here $z_1 = z - d$ is the distance from the S/F 
interface, and
\begin{eqnarray}
\Theta^{\mu}_{+} & = & 
B^{\mu}_{+} e^{\delta} + \alpha A^{\mu}_{+} e^{-\delta},
\nonumber \\
\Theta^{\mu}_{-} & = & 
\alpha A^{\mu}_{-} e^{\delta} + B^{\mu}_{-} e^{-\delta}, 
\nonumber \\
\Xi^{\mu}_{+} & = & 
\frac{ik_{-}}{p_3^{\downarrow}} B_{\mu}^{+} e^{\delta} + 
\alpha \frac{ik_{+}}{p_3^{\downarrow}} 
A^{\mu}_{+} e^{-\delta},
\nonumber \\
\Xi^{\mu}_{-} & = & - \left( 
\alpha \frac{ik_{+}}{p_3^{\downarrow}} A^{\mu}_{-} e^{\delta} + 
\frac{ik_{-}}{p_3^{\downarrow}} 
B^{\mu}_{-} e^{-\delta} \right).
\nonumber
\end{eqnarray}

Using Eqs.~(\ref{Four_a}), (\ref{F_ferro}), (\ref{F_index})
we get the expression for function $F$ averaged over the 
rapid oscillations 
\[
F(\omega, \xi, z=z') = 2m 
\sum_{\mu} \left[ 
\Phi_{\mu}^{-} b_{\mu\uparrow}^{+} + 
\Phi_{\mu}^{+} b_{\mu\uparrow}^{-} 
\right].
\]

 It follows from Eq.~(\ref{Phi}) for $\Phi_{\mu}^{\pm}$, that  
the dependence of function $F(\omega, \xi, z)$ on variable $z$ 
or $z_1 = z - d$ is given by a sum of the terms with
sine and cosine from arguments  $(p_3^{\uparrow} + p_3^{\downarrow})z_1$
and $(p_3^{\uparrow} - p_3^{\downarrow})z_1$. 
The terms with phases $(p_3^{\uparrow} + p_3^{\downarrow})z_1$
determine the short-periodic oscillations with respect to oscillations 
with larger period $\sim (p_3^{\uparrow} - p_3^{\downarrow})^{-1}$.
Neglecting the nonessential terms with short-periodic oscillations, 
the anomalous Green's function can be presented
in the form
\begin{eqnarray}
F(\omega, \xi, z=z') & = &
\widetilde \Theta(\omega,\xi) \cos(\Delta p_3 z_1) 
\label{F_oscill}
\\
& + & \widetilde \Xi(\omega,\xi) \sin(\Delta p_3 z_1),
\nonumber
\end{eqnarray}
where $z_1 = z - d$, 
$\Delta p_3 = p_3^{\uparrow} - p_3^{\downarrow}$, and 
$p_3^{\uparrow(\downarrow)}$ are given by Eqs.~(\ref{p1}), 
(\ref{p3}).

\begin{figure}[t]
\begin{center}
\includegraphics[scale = 0.9]{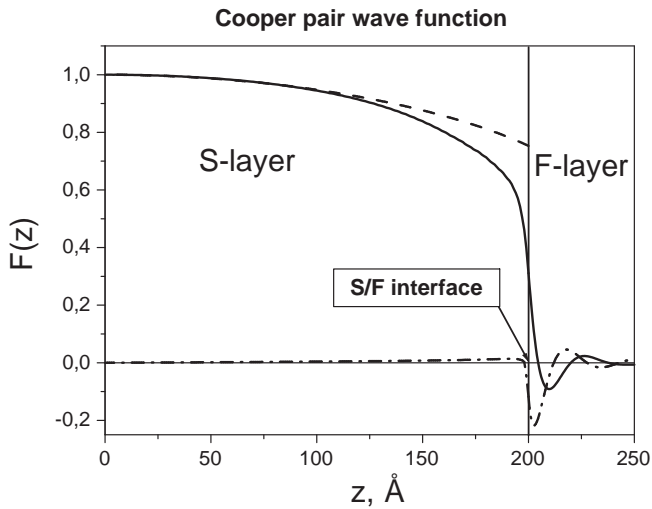}
\caption{The behavior of the anomalous Green's function $F(z)$ 
along the F/S/F structure with 
$d_S = 400$~\AA, $d_F = 50$~\AA, $\varepsilon_{\rm ex} = 1.156$~eV,
$l_{\uparrow} = 133$~\AA, $l_{\downarrow} = 35$~\AA.
The point $z = 200$~\AA\ is the S/F interface. Solid 
line~--- {\rm Re} $F(z)$, dashed dotted 
line~--- {\rm Im} $F(z)$. The contribution {\rm Re} $F_0(z)$ 
[see Eq.~(\ref{F0})] to the function {\rm Re} $F(z)$ in the S-layer 
is shown by dashed line.}
\end{center}
\end{figure}

 The real and imaginary parts of the function 
\[
F(z) = T \sum_{\omega}  \int_{0}^{\varepsilon_F}
d\xi\ F(\omega, \xi, z) 
\]
normalized on the value of its real part at the point 
$z = 0$ (the middle of the S-layer) are shown in Fig.~2.
The dashed line in Fig.~2
shows the contribution $F_0(z)$ [see Eq.~(\ref{F0})] to the 
function $F(z)$ in the S-layer. 
We can estimate the lengths 
responsible for the 
oscillations and decay using Eq.~(\ref{p3}) for momenta 
$p_3^{\uparrow}$ and $p_3^{\downarrow}$. Neglecting $\pm i\omega$ 
in Eq.~(\ref{p3}), since $|\omega|\le\omega_D$, we obtain
\[
p_3^{\uparrow(\downarrow)} = \sqrt{2m(\xi \pm h)}
\left[ 1 \pm \frac{i}{4}\tau^{-1}_{\downarrow(\uparrow)}
\frac{1}{\xi \pm h} + \dots\ \right]
\]
As far as the integration over $\xi$ goes from $0$ till $\varepsilon_F$,
then the damping of oscillations is determined by the value
\[
\sim \frac{1}{l_0} = \frac{1}{l_{\uparrow}} + 
\frac{1}{l_{\downarrow}}.
\]
Neglecting $\pm i \tau^{-1}_{\downarrow(\uparrow)}$ and $\pm i\omega$
in Eq.~(\ref{p3}), we get 
\[
p_3^{\uparrow(\downarrow)} = \sqrt{2m\xi}
\left[ 1 \pm \frac{h}{2\xi} + \dots\ \right].
\]
If $\xi \sim \varepsilon_F$, then 
\[
\Delta p_3 z_1 \sim \frac{\varepsilon_{\rm ex}}{v_F} z_1 = 
\frac{\pi z_1}{\xi_m^0},
\]
where $\xi_m^0 = \pi v_F/ \varepsilon_{\rm ex} = \pi \xi_F^0$ 
is the half-period of oscillations of $F(z)$ (the distance 
between the nearest zeros). Note, that $\xi_m^0 \ne \pi l_0$ in contrast
to what is found by using a quasiclassical approach.

The oscillatory terms in Eq.~(\ref{F_oscill}) arise 
due to quantum interference between two plane waves 
describing an electron and a hole propagating
in the ferromagnetic layer with different momenta 
$p_3^{\uparrow}$ and $- p_3^{\downarrow}$
along the $z$-axis. If $h\ne 0$
then $\Delta p_3 \ne 0$, and the oscillatory dependence 
of the Cooper pair wave function occurs due to the exchange 
field in the ferromagnet. If $h = 0$, then $F(z)$ exhibits only the 
exponential decay into the F-layer with characteristic 
length $l_0$. As it was already pointed out 
by many authors, the physical picture of the proximity
effect is similar to the nonuniform 
Fulde-Ferrel-Larkin-Ovchin\-nikov (FFLO) state \cite{FF,LO} 
which is characterized by the oscillatory dependent order parameter
and arises in a homogeneous superconductor in the presence of a strong 
enough uniform exchange field.

\section{Self-consistent evaluation of the 
order parameter}

In this section we proceed further by  
constructing the self-consistent solution of 
Eqs.~(\ref{Gor_F},\ref{Gor_S},\ref{self-cons}). 
In case of antiparallel orientation
of magnetizations in the ferromagnetic layers
the self-consistency can
 be reached if the order parameter 
$\Delta(z)$ takes the real values. 
We will search the 
self-consistent solution of 
Eq.~(\ref{self-cons}) 
in the S-layer assuming 
that in this equation the function $F(\omega,\xi,z)$ 
is replaced by its first contribution $F_0(\omega,\xi,z)$ 
given by Eq.~(\ref{F0}). Function $F_0(z)$ is shown by dashed line 
in Fig.~2 and can be approximated by a simple analytical function on $z$
like $\propto \cos(qz)$, where $q$ is a parameter. 
In order 
to take into account the correction  $F_1(z)$ 
one has to choose a more complex class of sample functions 
for $\Delta(z)$. However, this will not change the results 
significantly.

 Let us look for $\Delta(z)$ in the form
\[ 
\Delta(z) = \Delta \cos(qz) \simeq 
\Delta \left(1 - \frac{q^2z^2}{2} \right),
\]
where the wave vector $q$ (which has to be found) is small.
The magnitude $\Delta(d) = (1 - \delta_0)\Delta$ 
defines the amplitude of the superconducting 
order parameter at the S/F interface (see Fig.~2),  
here $\delta_0 = q^2 d^2/2 \sim 0.1$ is a small parameter. 
Following the well-known WKB-approximation \cite{WKB}, we 
search the solutions of Schr\"odinger's equation (\ref{Schrod}) 
in the form
\[
\psi(z) = \left( 
 e^{i\xi_+(z)} \atop e^{i\xi_-(z)} \right).
\]
For $\xi_{\pm}$ we get the system of equations
\begin{eqnarray}
\left[ i\omega + \xi + \frac{1}{2m}
\left( i\xi''_{+} - \xi_{+}'^{2} \right) \right] e^{i\xi_{+}} + 
\Delta(z) e^{i\xi_{-}} = 0, 
\label{xi_syst} \\
\Delta^{*}(z) e^{i\xi_{+}} + 
\left[ i\omega - \xi - \frac{1}{2m}
\left( i\xi''_{-} - \xi_{-}'^{2} \right) \right] e^{i\xi_{-}} = 0,
\nonumber
\end{eqnarray}
where $\xi = \varepsilon_F - \kappa^2/2m$ and primes 
above $\xi_{\pm}$ denote the derivatives by $z$.

In case of $q = 0$, $\Delta(z) = \Delta$,
four solutions of system (\ref{xi_syst}) are 
$\xi_{+}^0 = \pm k_{+} z$,
$\xi_{-}^0 = \pm  k_{+} z + i \log \alpha$ and 
$\xi_{+}^0 = \pm k_{-} z + i \log \alpha$,
$\xi_{-}^0 = \pm  k_{-} z $ which give four 
initial eigenfunctions of the 
non-perturbed Eq.~(\ref{Schrod}) 
($q = 0$):
\[
u_{\pm}^0(z) = \left( 1 \atop \alpha \right) e^{\pm ik_{+} z}, 
\quad 
v_{\pm}^0(z) = \left( \alpha \atop 1 \right) e^{\pm ik_{-} z}.
\]

Consider, for example, the perturbed solution $u_{+}(z)$ 
which corresponds to $u_{+}^0(z)$ in the case of $q \ne 0$. 
We look for the phases $\xi_{\pm}$ in the form 
$\xi_{\pm} = \xi_{\pm}^0 + \eta_{\pm}$, where 
$\xi_{+}^0 = k_{+} z$,
$\xi_{-}^0 = k_{+} z + i \log \alpha$ for $u_{+}^0(z)$. 
The typical order of $\eta_{\pm}$ is 
$ \sim q z \sim qd = \sqrt{2\delta_0} < 1$.
By linearizing the system (\ref{xi_syst}) with respect to 
$\eta_{\pm}$ we come to the following equations  
\begin{eqnarray}
- \frac{1}{\alpha} \frac{k_{+}}{ m}\ \eta_{+}' + 
i \Delta (\eta_{-} - \eta_{+} ) = \Delta \frac{q^2 z^2}{2},
\label{eta_syst} \\
\alpha \frac{k_{+}}{m}\ \eta_{-}' - 
i \Delta (\eta_{-} - \eta_{+} ) =  \Delta \frac{q^2 z^2}{2}.
\nonumber
\end{eqnarray}
We also have dropped the terms with $\eta_{\pm}^{\prime 2}$ 
and $\eta_{\pm}^{\prime\prime}$ which are small
as compared to $k_{\pm}\eta_{\pm}^{\prime}$, 
since $\eta_{\pm}^{\prime 2} \sim \eta_{\pm}^{\prime} \eta_{\pm}/d 
\sim \eta_{\pm}^{\prime} \sqrt{2\delta_0} /d \ll 
k_{\pm}\eta_{\pm}^{\prime}$ if $d \sim 200$~\AA\ 
and $k_{\pm} \sim 0.5$~\AA$^{-1}$, and 
$\eta_{\pm}^{\prime\prime} \sim \eta_{\pm}^{\prime}/d 
\ll k_{\pm}\eta_{\pm}^{\prime}$.  The equations
similar to Eqs.~(\ref{eta_syst}) can be written also 
for phases $\eta_{\pm}$ which determine 
other three solutions $u_{-}(z)$ and $v_{\pm}(z)$.
Solving these equations we get 
\[
u_{\pm}(z) = \left( 
{ \ e^{\pm i [k_{+}z + \eta_{+}^{(\pm)}(z)]} } \atop 
{ \alpha e^{\pm i[k_{+}z + \eta_{-}^{(\pm)}(z)]} } \right),
\]
with
\begin{eqnarray}
\eta_{\pm}^{(+)}(z) & = &  
\tau_3^{\pm} z^3 + \tau_2^{\pm} z^2 + \tau_1^{\pm} z + \tau_0^{\pm},
\nonumber \\
\eta_{\pm}^{(-)}(z) & = &  
\tau_3^{\pm} z^3 - \tau_2^{\pm} z^2 + \tau_1^{\pm} z - \tau_0^{\pm};
\nonumber 
\end{eqnarray}
and
\[
v_{\pm}(z) = \left( 
{ \alpha e^{\pm i [k_{-}z + \zeta_{+}^{(\pm)}(z)]} } \atop 
{ \ e^{\pm i [ k_{-}z + \zeta_{-}^{(\pm)}(z)]} } \right),
\]
where
\begin{eqnarray}
\zeta_{\pm}^{(+)}(z) & = &  
\rho_3^{\pm} z^3 + \rho_2^{\pm} z^2 + \rho_1^{\pm} z + \rho_0^{\pm},
\nonumber \\
\zeta_{\pm}^{(-)}(z) & = &  
\rho_3^{\pm} z^3 - \rho_2^{\pm} z^2 + \rho_1^{\pm} z - \rho_0^{\pm}.
\nonumber 
\end{eqnarray}
The expressions for coefficients $\tau^{\pm}_i$ and $\rho^{\pm}_i$
of polynomials are given in Appendix B.

 Next the procedure of evaluation of the anomalous Green's 
function in the S-layer is similar to one described in 
details in Secs.~II and III for the case of $\Delta(z) = \Delta$.
By representing the solutions $\varphi_{\mu}(z)$
and $\psi_{\mu}(z)$ of Eq.~(\ref{Schrod}) as a linear 
combination of eigenfunctions $u_{\pm}(z)$ and $v_{\pm}(z)$
similar to representation 
(\ref{sol_S}), we can find the new coefficients $A^{\mu}_{\pm}$, 
$B^{\mu}_{\pm}$, $C^{\mu}_{\pm}$, $D^{\mu}_{\pm}$ solving
the system of 4 linear equations. By evaluating the currents
$j_{\mu\rho}$ at the point $z = 0$ ($j_{\mu\rho}$ do not depend on $z$), 
we obtain the expressions for $j_{\mu\rho}$ similar to Eq.~(\ref{j_sc})
where $k_{\pm}$ should be replaced by 
\[
\widetilde k_{+} = k_{+} + 
\frac{\tau_1^+ - \alpha^2 \tau_1^-}{1 - \alpha^2},
\quad
\widetilde k_{-} = k_{-} + 
\frac{\rho_1^- - \alpha^2 \rho_1^+}{1 - \alpha^2},
\]
and $\theta_{\pm} = 2ik_{\pm} d = \theta \pm i\delta$. 
The substitutions $k_{\pm} \to \widetilde k_{\pm}$ also 
have to be made in Eq.~(\ref{a_coeff}) for $a_{\mu\rho}$
and in the expression for $\mathrm{det}J$ (see Appendix A).
Finally, the anomalous Green's function 
$F(\omega, \xi, z)$ is given by 
Eqs.~(\ref{F0}) where $\Lambda_{\pm}^{\mu}$ and
$\Sigma_{\pm}^{\rho}$ are replaced by 
new functions $\widetilde \Lambda_{\pm}^{\mu}$ and
$\widetilde \Sigma_{\pm}^{\rho}$: 
\begin{eqnarray}
\widetilde \Lambda^{\mu}_{+} & = & 
\alpha A^{\mu}_{+} e^{-\delta_1 + i\eta_{-}^{(+)}} 
 +  B^{\mu}_{+} e^{\delta_1 + i\zeta_{-}^{(+)}}, 
\nonumber  \\
\widetilde \Lambda^{\mu}_{-} & = & 
\alpha A^{\mu}_{-} e^{\delta_1 - i\eta_{-}^{(-)}} 
 +  B^{\mu}_{-} e^{-\delta_1 - i\zeta_{-}^{(-)}}, 
\nonumber
 \\
\widetilde \Sigma^{\rho}_{+} & = & 
C^{\rho}_{-} e^{-\delta_2 - i\eta_{+}^{(-)}} 
 + \alpha D^{\rho}_{-} e^{\delta_2 - i\zeta_{+}^{(-)}}, 
\nonumber \\
\widetilde \Sigma^{\rho}_{-} & = & 
C^{\rho}_{+} e^{ \delta_2 + i\eta_{+}^{(+)}} 
 + \alpha D^{\rho}_{+} e^{-\delta_2 + i\zeta_{+}^{(+)}}, 
\nonumber
\end{eqnarray}
here $\delta_1$, $\delta_2$, $\eta_{\pm}^{(\pm)}$ and 
$\zeta_{\pm}^{(\pm)}$ are functions on $z$.
 The fixed 
point $q = q_*$ which determines the  
order parameter $\Delta(z)$
 has to be found 
numerically by solving Eq.~(\ref{self-cons})
using the iterative procedure.

\section{Critical temperature $T_c$}

If the anomalous Green's function in the S-region
$F(\omega, \xi, z) = F(\omega, \xi, z = 0) \cos (q_* z)$ 
is known, the superconducting transition temperature $T_c$ 
can be found. Up to now we assume the "clean" 
limit for a superconductor. The corrections (\ref{sc_S}) due to scattering
will be taking into account further. Let us introduce the function 
\[
F_{\omega} = \frac{1}{k_F} \int_{0}^{\varepsilon_F} d\xi\
F(\omega, \xi,z=0),
\]
where $k_F$ is Fermi momentum in the S-layer. 
This integral can be evaluated only numerically. 
However, we can approximate $F_{\omega}$ by the
analytical function of argument~$\omega$.
Let us represent $F_{\omega}$ in the form
\[
F_{\omega} = 
\frac{\Delta}{\sqrt{\omega^2 + \Delta^2 }}\
F^{(1)}_{\omega}.
\]
For the bulk superconductor $F^{(1)}_{\omega} = 1$.
Let $T \to T_c$, therefore $\Delta \ll \Delta(0)$, 
where $\Delta(0)$ is the order parameter at $T=0$.
If $\omega$ takes values from $0$ till $\sim 5\omega_D$, 
$F^{(1)}_{\omega}$ can be well approximated by the
following function 
\begin{equation}
F^{(1)}_{\omega} \simeq
A_0 \tanh \left( \frac{\gamma_0|\omega|}{2\omega_D}
\right).
\label{F1_appr}
\end{equation}
The coefficients $A_0$ and $\gamma_0$ are found 
numerically by minimizing the norm of a difference between 
the exact and approximate function. 
These coefficients are non-monotonic functions of the 
F-layer thickness $d_F$ when $d_S$ is fixed.
For typical values of the parameters describing the 
F/S/F structure 
the magnitudes of $A_0$ and $\gamma_0$ are
$A_0 \sim 0.9$ and $\gamma_0 \sim 4.0$.

 The scattering in the S-layer is introduced by 
Eq.~(\ref{sc_S}). Numerical analysis shows 
that in Eq.~(\ref{sc_S}) the Green's function 
$G_{\uparrow\uparrow}^{ss}(z,z)$ does not depend on $z$
in the S-region and its real part is negligibly small.
Obviously, $\Delta_{\omega}(z) = \Delta_{\omega} \cos (q_{*}z)$.
From numerical analysis it follows that
$G_{\omega}$ in the S-layer
 can be represented as
\begin{eqnarray}
G_{\omega} & = & \frac{1}{k_F} \int_{0}^{\varepsilon_F} 
d \xi\ G_{\uparrow\uparrow}^{ss}(\omega, \xi, z = 0)
\nonumber \\
& \approx & - A_0 \frac{i \omega}{\sqrt{\omega^{2} + \Delta^2 }}.
\label{Gw_appr}
\end{eqnarray}
Taking into account Eqs.~(\ref{F1_appr}) and (\ref{Gw_appr}), 
equations (\ref{sc_S}) can be written in the form 
similar to the case of bulk superconductor \cite{Svidz}:
\begin{eqnarray}
\omega' & \approx & \omega + \frac{A_0}{2\tau_0} 
\ \frac{\omega'}{\sqrt{\omega^{'2} + \Delta_{\omega}^2 }},
\nonumber \\
\Delta_{\omega} & \approx & \Delta + \frac{A_0}{2\tau_0} 
\ \frac{\Delta_{\omega}}{\sqrt{\omega^{'2} + \Delta_{\omega}^2 }},
\label{wd}
\end{eqnarray}
where $\tau_0^{-1} = 2\pi c u_0^2 N(\varepsilon_F)$ is the inverse 
life-time of quasiparticles in the superconductor, and 
$N(\varepsilon_F) = m k_F/2\pi^2$ is density of states at the Fermi
energy. Deriving Eq.~(\ref{wd}) we took into account that, 
if $\frac12 \tau_0^{-1} \sim \omega_D \sim 300$~K
corresponding to mean free path $l_s \sim 130$~\AA, then 
for $\omega' \simeq  \omega + {A_0}/{2\tau_0}$ and $\gamma_0 \sim 4.0$
we have $\tanh \left( \gamma_0 |\omega'|/2 \omega_D \right) \approx 1$.

  Equations (\ref{wd}) can be written as \cite{Svidz}
\begin{eqnarray}
\omega' = \omega \eta(\omega), \quad 
\Delta_{\omega} = \Delta \eta(\omega),
\nonumber \\
\eta(\omega) = 1 +  
\frac{A_0}{2\tau_0 \sqrt{\omega^{2} + \Delta^2 }}.
\label{etaw}
\end{eqnarray}
Using (\ref{etaw}) and (\ref{self-cons})
we come to the equation for $T_c$:
\begin{equation}
\pi\rho T_c \sum_{\omega} \frac{1}{|\omega|}
\tanh \left( \frac{\gamma_0|\omega'|}{2\omega_D}
\right) = 1, 
\label{Tc_w}
\end{equation}
where
\[
\omega' = \omega + \frac{A_0}{2\tau_0}, \quad
\rho = \rho_0 A_0 < \rho_0,
\]
and $\rho_0 = \lambda N(\varepsilon_F)$ is the renormalized 
coupling constant. By carrying out the summation over
Matsubara frequencies 
$\omega = \pi T_c (2n + 1)$ in Eq.~(\ref{Tc_w}), 
we get the equation 
for reduced critical temperature $\tau = T_c/T_c^0$:
\begin{equation}
\tau = \exp \left\{ 
\left( \frac{1}{\rho_0} - \frac{1}{\rho} \right) -
\Phi \Bigl [\eta_0(\tau)\Bigr ] 
\right\},
\label{eq_Tc}
\end{equation}
where
\[
\Phi(\eta_0) = \sum_{n=0}^{+\infty}
\frac{4\Gamma_0 e^{-(2n+1)\eta_0}}
{(2n+1)(1 + \Gamma_0 e^{-(2n+1)\eta_0 } )},
\]
\[
\eta_0(\tau) = \frac{\gamma_0\pi T_0}{\omega_D}\tau, \quad
\tau = T_c/T_c^0,
\]
\[ 
\Gamma_0 = \exp \left( -\frac{\gamma_0 A_0}{2\tau_0\omega_D} \right),
\]
and $T_c^0 = 2 \pi^{-1}\omega_D \gamma e^{-1/\rho_0}$ 
($\gamma = e^C,\ C = 0.577 \dots \ $) 
is transition temperature of the bulk superconductor.
 
\section{Results and discussion}

In this section we present the results of numerical 
calculation of the critical temperature $T_c$. We first 
focus on the general features of a behavior of the 
system. Next we consider selected experimental data 
which 
can be interpreted in the framework of the given model.

\subsection{Oscillatory behavior of $T_c$}

\begin{figure}[t]
\begin{center}
\includegraphics[scale = 0.35]{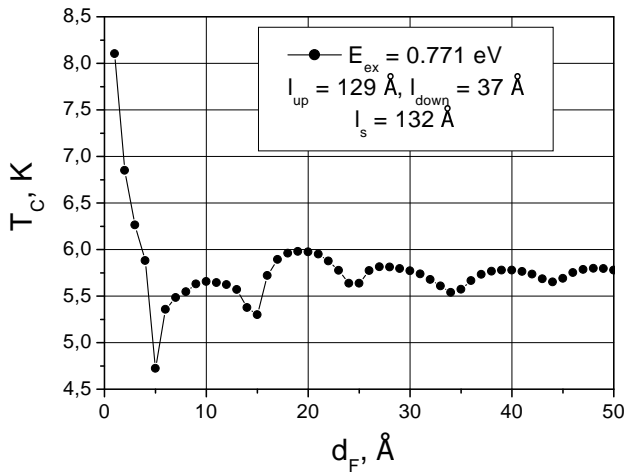}
\caption{Critical temperature $T_c$ for the F/S/F 
trilayer as a function of the ferromagnetic layer thickness $d_F$. 
The parameters are
$\varepsilon_{\rm ex} = 0.771$~eV (exchange field in the ferromagnet), 
$l_{\uparrow} = 129$~\AA,
 $l_{\downarrow} = 37$~\AA, 
$l_s  = 132$~\AA (mean free paths in the F and S layers),
$k_F  = 0.826$~\AA$^{-1}$
(Fermi momentum in the superconductor), 
$\omega_D = 275$~K, $T_c^0 = 9.25$~K. 
}
\end{center}
\end{figure}

\begin{figure}[b]
\begin{center}
\includegraphics[scale = 0.35]{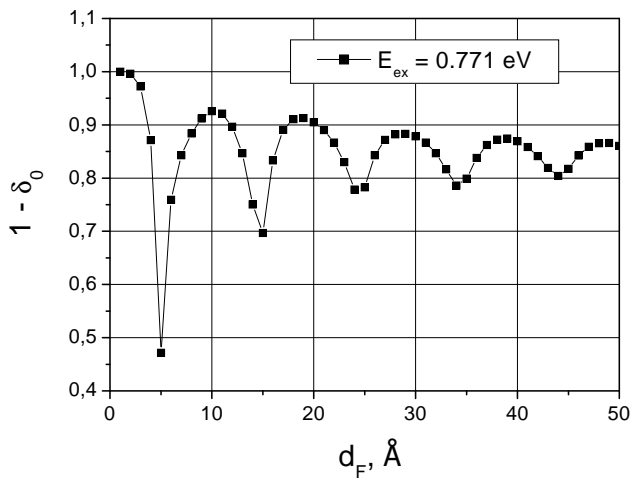}
\caption{The dependence of the normalized amplitude $(1-\delta_0)$ 
of the superconducting order parameter 
at the S/F interface as a function of
the F-layer thickness $d_F$ for the same parameters as in Fig.~3.}
\end{center}
\end{figure}

The typical dependence of critical temperature $T_c(d_F)$ 
with respect to ferromagnetic layer thickness $d_F$ with $d_S = 400$~\AA\
is shown in Fig.~3 where the model parameters are given
in the figure caption.
The effective electron mass is $m = m_e$ ($m_e$ is a 
bare electron mass). For superconductor we took $\omega_D = 276$~K 
and $T_c^0 = 9.25$~K which are the parameters of bulk Nb. 
The corresponding normalized magnitude $(1-\delta_0)$ of the 
order parameter at the S/F interface
as a functions of $d_F$ is shown in Fig.~4.

Both functions $(1-\delta_0)$ and $T_c(d_F)$ show the 
pronounced damping oscillatory behavior with the same period. 
The oscillatory behavior of $T_c(d_F)$ is a consequence 
of oscillations of the amplitude $\Delta(d) = \Delta(1 - \delta_0)$ of the 
order parameter at the S/F interface when $d_F$ is varying. 
The minima of $\Delta(d)$ 
correspond to minima of $T_c$ and the maxima of $\Delta(d)$ 
correspond to the maxima of $T_c$, as they should.
The oscillations of $\Delta(d)$ in turn are caused by the 
oscillations of the anomalous Green's function $F(z)$ in the F-layer.  
Function $F(z)$ must satisfy the zero boundary condition 
at the ferromagnet/vacuum interface. Because of oscillations 
of $F(z)$ in the F-region, the order 
parameter at the S/F interface is forced to adjust in such a way that the 
condition $F(a) = 0$ is fulfilled at the outer boundary $z = a$
of the F-layer.

\begin{table}[b]
\caption{The comparison of the periods of oscillations
$\xi_F$ predicted by formula (\ref{per_xi_F}) with the period 
$\widetilde \xi_F$ obtained from numerical analysis
for different values of exchange field $\varepsilon_{\rm ex}$, 
effective electron mass $m$ 
and a superconductor layer thickness $d_S$.
Other model parameters are the same as in Fig.~3. The accuracy
of determining of $\widetilde \xi_F$ is restricted 
by a finite step for $d_F$ in numerical calculations. }
\begin{ruledtabular}
\begin{tabular}{cccrr}
$\varepsilon_{\rm ex}$\ (eV) & $m\ (m_e)$ &
$d_S$\ (\AA) & $\xi_F$\ (\AA) & $\widetilde \xi_F$\ (\AA)\\ 
\hline
0.385 & 1.0 & 400 & 13.97 & 14.0 \\ 
0.771 & 1.0 & 400 & 9.98 & 10.0 \\ 
1.156 & 1.0 & 400 & 8.06 & 8.0 \\ 
2.027 & 1.0 & 400 & 6.09 & 6.0 \\ 
0.610 & 0.45 & 600 & 16.55 & 16.5 \\ 
\end{tabular}
\end{ruledtabular}
\end{table}

 The results of numerical analysis, presented in Table~I for 
different values of exchange field $\varepsilon_{\rm ex}$
and effective electron mass $m$, show that the period $\xi_F$
of $T_c$-oscillations is defined as
\begin{equation}
\xi_F = \frac{\pi}{\sqrt{m\varepsilon_{\rm ex} }}
= \sqrt{\pi \xi_m^0 k_F^{-1}}, 
\label{per_xi_F}
\end{equation}
here $k_F$ is the Fermi momentum in a superconductor. The period $\xi_F$, 
therefore, does not depend on the electron mean free paths in 
the S- and F-layers. The first minimum of 
$T_c(d_F)$ occurs at the thickness $\xi_F/2$, 
while the location of first maximum is $\xi_F$.

As can be seen from Fig.~5, 
the strong scattering in the ferromagnetic layers 
significantly damps the oscillations of $T_c$, but
their period remains unchanged 
for any values of the mean free paths $l_{\uparrow}$ and 
$l_{\downarrow}$. As it follows from the analysis presented in  Sec.~III, 
the reason of such a behavior is that 
the strong scattering in the F-region affects
only the length $l_0$ of decay of the Cooper 
pair wave function $F(z)$ but not the period $\sim \xi_m^0$ of its
oscillations. The less pronounced are the 
oscillations of $F(z)$ with respect to $z_1 = z - d$ 
in case of strong electron scattering, the less 
is the amplitude of oscillations of $\Delta(d)$ and $T_c$ 
with respect to the ferromagnetic layer thickness $d_F$. 
In case of extremely strong scattering,  
the coherent coupling
which was established due to these oscillations 
between two boundaries of ferromagnetic layer is destroyed and thus 
the oscillations of $T_c$ are suppressed completely.

We also observed that strong scattering in the S layer
(small mean free path $l_s$) suppresses the amplitude of $T_c$
oscillations (look at Fig.~6). 
The critical temperature is higher 
for smaller values of $l_s$
. The reason for it is that 
in the thin superconducting films $T_c$ is reduced 
with respect to $T_c^0$ due to dimensional 
effect, and the magnitude of $T_c$ depends on $d_S$ only 
via the dimensionless thickness $d_S/\xi_S$, 
where $\xi_S \propto \sqrt{\xi_0 l_s }$ is 
a coherence length for the dirty superconductor, $\xi_0$  
is a BCS coherence length. Small mean free path $l_s$, therefore,  
corresponds to large value of the effective film thickness $d_S/\xi_S$.

\subsection{Comparison with experiment}

\begin{figure}[t]
\begin{center}
\includegraphics[scale = 0.35]{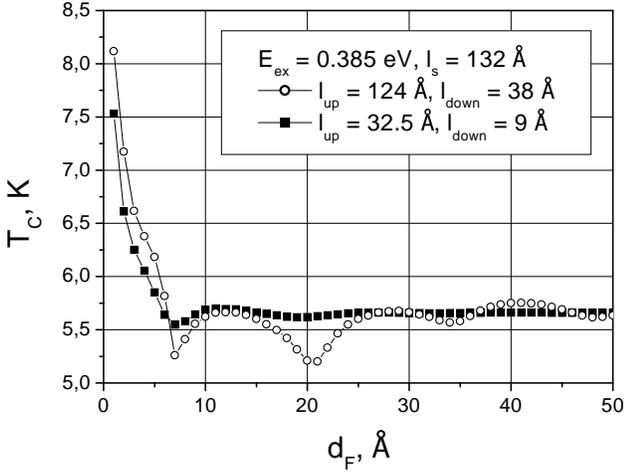}
\caption{The critical temperature $T_c(d_F)$ for 
the case of weak and strong scattering in 
the ferromagnetic layers; $\varepsilon_{\rm ex} = 0.385$~eV,
$l_s = 132$~\AA. Dots (weak scattering): $l_{\uparrow} = 124$~\AA, 
$l_{\downarrow} = 38$~\AA; squares 
(strong scattering): $l_{\uparrow} = 32.5$~\AA, 
$l_{\downarrow} = 9$~\AA.}
\end{center}
\end{figure}

\begin{figure}[b]
\begin{center}
\includegraphics[scale = 0.35]{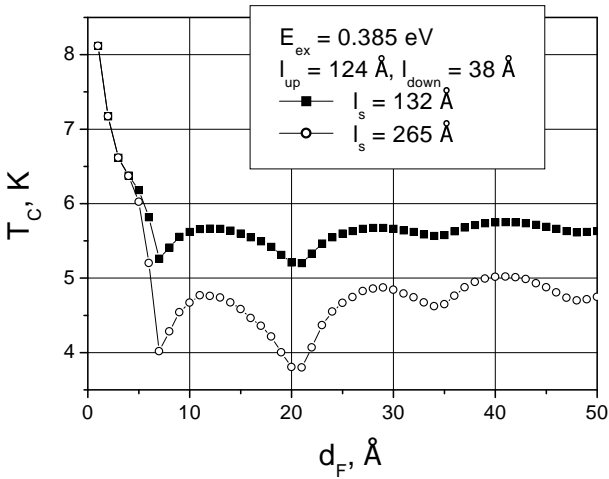}
\caption{The critical temperature $T_c(d_F)$ for 
different values of the mean free paths in the S-layer;
$\varepsilon_{\rm ex} = 0.385$~eV,
$l_{\uparrow} = 124$~\AA, $l_{\downarrow} = 38$~\AA. 
Squares: $l_s = 132$~\AA; dots: $l_s = 265$~\AA.
}
\end{center}
\end{figure}

Experimental situation on the oscillatory behavior of $T_c(d_F)$ 
in the S/F structures is known to be controversial. 
Nevertheless, there are two groups of 
experiments described in the literature
where oscillations of $T_c(d_F)$ were clearly 
observed and the 3$d$ ferromagnets 
were used as F layers --- these are reports on Fe/Nb/Fe trilayers
by M\"uhge {\it et al.}\cite{Muhge}, and Nb/Co and V/Co 
multilayers by Obi {\it et al.}\cite{Obi}.

 In Fig.~7 the fitting is shown to experimental data by M\"uhge {\it et al.}
\cite{Muhge} for Fe($d_F$)/Nb(400\,\AA)/Fe($d_F$) trilayers prepared
by rf sputtering. According to formula (\ref{per_xi_F}) the period
$\xi_F$ of oscillations is determined by exchange splitting energy in the
ferromagnet. If we take the value $\varepsilon_{\rm ex}^d 
\simeq 0.149~{\rm Ry} = 2.03$~eV (Ref.~\onlinecite{Moruzzi}) 
of exchange splitting of the Fe $d$-bands 
near the Fermi energy and put $m = m_e$, 
we obtain $\xi_F^{(d)} = 6.09$\,\AA\
(see Table~I) which is too small as compared to 
the location of a maximum at 
$d_F \sim 10\div 15$\,\AA\ in Fig.~7. 
However, we can assume that in the S and F layers 
the Cooper pairs are formed by $s$ electrons of Nb and Fe. 
The value of exchange splitting 
$\varepsilon_{\rm ex}^s = 0.028~{\rm Ry} = 0.381$~eV
at the bottom of Fe $s$ bands 
($\Gamma$ point, Ref.~\onlinecite{Moruzzi}
) 
together with $m = m_e$ gives the period $\xi_F = 14.05$~\AA. 
Thus, the first minimum of $T_c(d_F)$ 
is at the point $\xi_F/2 \approx 7$~\AA, and the first 
maximum is at $\xi_F \approx 14$~\AA. From Fig.~7 it follows that 
these values correlate with positions of minimum and maximum 
of $T_c$
which can be roughly determined from the scattered experimental points. 
We have put $\varepsilon_F = 0.387$~Ry corresponding 
to the $s$ band of Nb (Ref.~\onlinecite{Moruzzi}) 
which gives the Fermi momentum value $k_F = 1.18$~\AA$^{-1}$ for 
$m = m_e$. We used $\omega_D = 276$~K and $T_c^0 = 9.25$~K for Nb.
The fitting parameters are the values of mean free paths 
in Fe and Nb which were estimated approximately
as $l_{\uparrow} = 120$~\AA, 
$l_{\downarrow} = 40$~\AA, and $l_s = 269$~\AA.
Note, that magnetic measurements by M\"uhge {\it et al.} 
showed that thin Fe layers were not magnetic for $d_F \le 7$~\AA, 
and it was assumed that magnetically "dead" Fe-Nb alloy 
of a thickness about 7~\AA\ was formed at the interfacial S/F region 
for all samples with different $d_F$. 
M\"uhge {\it et al.} qualitatively explained the 
observed non-monotonic behavior of $T_c(d_F)$ in terms of a rather 
complex behavior of this magnetically "dead" Fe-Nb layer when $d_F$ 
was varying (see details in Ref.~\onlinecite{Muhge}). 
They also argued that a non-monotonic $T_c(d_F)$
behavior in their case could not be possible due to the mechanism of 
$\pi$ coupling as it was predicted for the S/F multilayers
because of a single S layer in the trilayer system. 
Indeed, the well-known theoretical prediction by Buzdin 
{\it et al.} \cite{Buzdin_JETP,Buzdin_PRB} 
ascribes the oscillatory behavior of $T_c(d_F)$
to the periodical switching of the ground state energy
between 0- and $\pi$-phases of the order parameter 
if the neighboring S-layers in the S/F multilayer 
are coupled. However, it follows from the above analysis 
that the oscillatory behavior of $T_c(d_F)$ 
does not necessary require the $\pi$ coupling 
and can occur also for a trilayer (or bilayer) F/S/F structure.

\begin{figure}[t]
\begin{center}
\includegraphics[scale = 0.34]{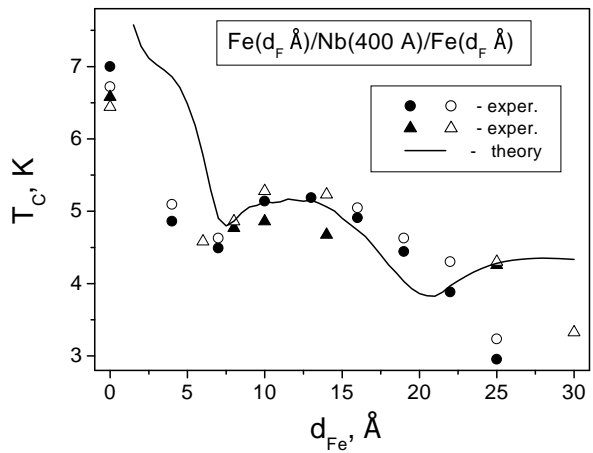}
\caption{The comparison of the theoretical $T_c(d_F)$ curve
with experiment by M\"uhge {\it et al.} 
(Ref.~\onlinecite{Muhge}
) for Fe/Nb(400\AA)/Fe trilayers. 
The fitting parameters are
$l_{\uparrow} = 120$~\AA, $l_{\downarrow} = 40$~\AA, $l_s  = 269$~\AA.}
\end{center}
\end{figure}

\begin{figure}[b]
\begin{center}
\includegraphics[scale = 0.35]{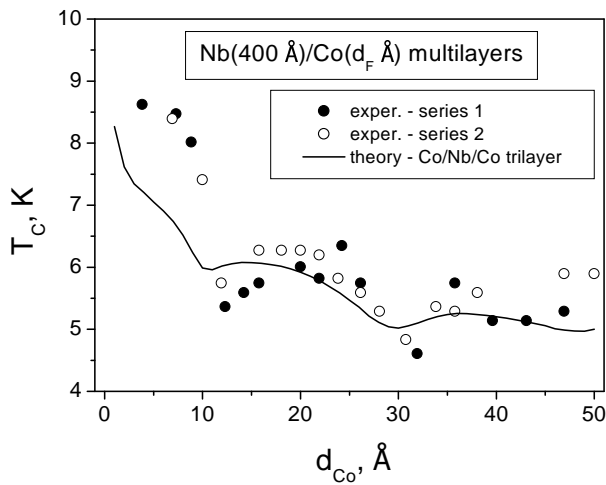}
\caption{The comparison of the theoretical $T_c(d_F)$ curve
with experiment by Obi {\it et al.} 
(Ref.~\onlinecite{Obi}) for Nb(400\AA)/Co multilayers. 
The fitting parameters are $l_{\uparrow} = 240$~\AA, 
$l_{\downarrow} = 80$~\AA, $l_s  = 188$~\AA.}
\end{center}
\end{figure}

Let us consider the 
experiments on Nb/Co multilayers 
by Obi {\it et al.}~\cite{Obi}.
The theoretical curve $T_c(d_F)$ in comparsion with experimental 
data is shown 
in Fig.~8. The exchange splitting of 
Co spin-up and down $s$ bands at $\Gamma$ 
point is $\varepsilon_{\rm ex} = 0.014$~Ry 
(Ref.~\onlinecite{Moruzzi}) 
which gives $\xi_F = 19.87$~\AA\ ($m = m_e$).
The first and second minimum of $T_c(d_F)$ should,  
therefore, be placed at points $d^1_{\rm min} = 10$~\AA\ and
$d^2_{\rm min} = 30$~\AA. These values correlate with 
values 12~\AA\ and 32~\AA\ obtained from experiment. 
The fitting mean free paths are
$l_{\uparrow} = 240$~\AA, $l_{\downarrow} = 80$~\AA, 
and $l_s = 188$~\AA. We have to note that in experiment
Nb/Co structures are multilayers. 
A qualitative resemblance of theoretical 
$T_c$ curve calculated for a trilayer structure with experimental 
points for a multilayer and the agreement between theoretical and 
experimental values of $d^1_{\rm min}$ and $d^2_{\rm min}$ 
allows us to assume that neighboring S layers were 
decoupled in the experiment. 
As it was observed by Strunk {\it et al.} \cite{Strunk} 
for similar Nb/Fe multilayered system 
(where F-layer is 3$d$ transition metal), the decoupling
regime is set when $d_F$ is larger than some
critical value $d_c^{\rm dc}$ which in turn is less than 
the critical thickness $d_c^F$ of the onset of ferromagnetism. 
This threshold value was $d_c^F \approx 7$~\AA\ in experiments by 
Obi {\it et al.} \cite{Obi} 
In Ref.~\onlinecite{Obi} it was noted that $d_c^F$ was less
than the first minimum of $T_c$ at $d^1_{\rm min} \approx 12$~\AA, 
so that for Nb/Co system the first
minimum could not be ascribed to the onset of ferromagnetism
as it was argued by M\"uhge {\it et al.} 
for the Fe/Nb/Fe system\cite{Muhge}. Our theoretical explanation
assuming the decoupling regime is incorrect only for very thin Fe layers
with $d_F < d_c^{\rm dc}$ when, probably, 
the Fe films are nonmagnetic due to alloying effect.

 Note also, that experiments by Obi {\it at el.} \cite{Obi}
on Nb$_{1-x}$Ti$_x$/Co multilayers with Nb$_{1-x}$Ti$_x$ 
alloy being superconductor with small coherence length 
did not reveal the oscillatory behavior of $T_c$
but showed only a small reduction of the critical temperature 
$T_c \approx 8$~K for large $d_F$ as compared to 
the bulk value $T_c^0 \approx 9.2$~K 
(see Fig.~3 in Ref.~\onlinecite{Obi}). 
Therefore, the observation of increasing of 
$T_c$ when the scattering is strong in S-layer together with damping
of oscillations for small $l_s$ (see Fig.~6) is 
in a qualitative 
agreement with these experimental observations. 

\section{Summary}

In conclusion, we have presented a theory of proximity effect
in F/S/F trilayer nanostructures where S is a superconductor
and F are layers of 3$d$ transition ferromagnetic metal. As a starting point 
of our calculations, we took the system of Gor'kov equations, which determine 
the normal and anomalous Green's functions. The solution of these equations 
was found together with a self-consistent evaluation 
of the superconductor order parameter. In accordance with the known
quasiclassical theories of proximity effect for S/F multilayers
\cite{Buzdin_JETP,Buzdin_PRB,Demler,Khusainov},
we found that due to a presence of exchange field in the ferromagnet
the anomalous Green's function $F(z)$ exhibits 
damping oscillations in the F-layer
as a function of a distance $z$ from the S/F interface. In the presented 
model a half-period of oscillations of $F(z)$
is determined by the length $\xi_m^0 = \pi v_F/\varepsilon_{\rm ex}$, 
where $v_F$ is the Fermi velocity, $\varepsilon_{\rm ex}$ is 
the exchange 
field, and the length of damping is 
given by $l_0 = (1/l_{\uparrow} + 1/l_{\downarrow})^{-1}$, 
where $l_{\uparrow}$ 
and $l_{\downarrow}$ are spin-dependent 
mean free paths in the
 ferromagnetic layer. 
The oscillations of the anomalous
Green's function (Cooper pair wave function) in the F-region 
and a zero boundary condition at the ferromagnet/vacuum interface
give rise to the oscillatory dependence of the superconductor
order parameter at the S/F interface {\it vs} the F-layer 
thickness~$d_F$. These oscillations result in oscillations
of the superconductor transition temperature $T_c(d_F)$
with a period $\xi_F = \pi/\sqrt{m \varepsilon_{\rm ex}}$. 
Thus we have demonstrated that the nonmonotonic oscillatory 
dependence of critical temperature $T_c(d_F)$ does not necessarily
require the mechanism of $\pi$-coupling between neighboring 
superconducting layers as it takes place in the S/F 
multilayers \cite{Buzdin_JETP,Buzdin_PRB}.
The strong electron scattering either in the superconductor 
or in the ferromagnet significantly suppresses the oscillations.
In case of extremely strong scattering in the ferromagnet
the length of damping $l_0$ becomes very short and
the oscillations of $T_c$ are suppressed completely. 
The reason of that is the loss of coherent "coupling" 
between two boundaries of ferromagnetic layer that was established 
due to oscillations of Cooper pair wave function $F(z)$. 
We compared our results with existing data
on $T_c(d_F)$ for Fe/Nb/Fe trilayers \cite{Muhge}
and V/Co multilayers \cite{Obi}, where F's are 3$d$ ferromagnets,
and found reasonable agreement with theory and experiment.

\section*{Acknowledgments}

This work was partially supported by Russian Foundation for 
Basic Research under grant No.~01-02-17378. A.B.\ is grateful to NATO
Science Fellowships Program for a financial support.

\appendix

\section{}
The determinant 
$\mathrm{Den} = \mathrm{det} J/(1-\alpha^2)^2$ 
of the matrix of currents (Eq.~\ref{currents})
is given by the expression 
\[
\mathrm{Den} = - D_0 + 
\Gamma_{+} e^{2i\theta} + \Gamma_{-} e^{-2i\theta}, 
\]
where
\begin{eqnarray}
D_0 & = & 4k_{+}^2
\left|
\begin{array}{cc}
A^{\uparrow}_{+} & A^{\downarrow}_{+} \\
A^{\uparrow}_{-} & A^{\downarrow}_{-}
\end{array}
\right| \times
\left|
\begin{array}{cc}
C^{\uparrow}_{+} & C^{\downarrow}_{+} \\
C^{\uparrow}_{-} & C^{\downarrow}_{-}
\end{array}
\right|
\nonumber \\
& + & 4k_{-}^2
\left|
\begin{array}{cc}
B^{\uparrow}_{+} & B^{\downarrow}_{+} \\
B^{\uparrow}_{-} & B^{\downarrow}_{-}
\end{array}
\right| \times
\left|
\begin{array}{cc}
D^{\uparrow}_{+} & D^{\downarrow}_{+} \\
D^{\uparrow}_{-} & D^{\downarrow}_{-}
\end{array}
\right| 
\nonumber \\
& + & 4k_{+}k_{-} e^{2\delta}
\left|
\begin{array}{cc}
A^{\uparrow}_{-} & A^{\downarrow}_{-} \\
B^{\uparrow}_{+} & B^{\downarrow}_{+}
\end{array}
\right| \times
\left|
\begin{array}{cc}
C^{\uparrow}_{+} & C^{\downarrow}_{+} \\
D^{\uparrow}_{-} & D^{\downarrow}_{-}
\end{array} 
\right|
\nonumber \\
& + & 4k_{+}k_{-} e^{-2\delta}
\left|
\begin{array}{cc}
A^{\uparrow}_{+} & A^{\downarrow}_{+} \\
B^{\uparrow}_{-} & B^{\downarrow}_{-}
\end{array}
\right| \times
\left|
\begin{array}{cc}
C^{\uparrow}_{-} & C^{\downarrow}_{-} \\
D^{\uparrow}_{+} & D^{\downarrow}_{+}
\end{array} 
\right|, 
\nonumber  \\ \medskip
\Gamma_{+} & = & 4k_{+}k_{-}
\left|
\begin{array}{cc}
A^{\uparrow}_{+} & A^{\downarrow}_{+} \\
B^{\uparrow}_{+} & B^{\downarrow}_{+}
\end{array}
\right| \times
\left|
\begin{array}{cc}
C^{\uparrow}_{-} & C^{\downarrow}_{-} \\
D^{\uparrow}_{-} & D^{\downarrow}_{-}
\end{array} 
\right|, 
\nonumber \medskip \\
\Gamma_{-} & = & 4k_{+}k_{-}
\left| 
\begin{array}{cc}
A^{\uparrow}_{-} & A^{\downarrow}_{-} \\
B^{\uparrow}_{-} & B^{\downarrow}_{-}
\end{array}
\right| \times
\left|
\begin{array}{cc}
C^{\uparrow}_{+} & C^{\downarrow}_{+} \\
D^{\uparrow}_{+} & D^{\downarrow}_{+}
\end{array} 
\right|, 
\nonumber
\end{eqnarray}
and $A^{\mu}_{\pm}$, $B^{\mu}_{\pm}$, $C^{\mu}_{\pm}$, $D^{\mu}_{\pm}$ 
are coefficients introduced in Eq.~(\ref{sol_S}).

\section{}

Let us define the quantities
\begin{eqnarray}
\lambda^{-1}_{\pm} & = & \frac{2m}{k_{\pm}} \sqrt{\omega^2 + \Delta^2},
\nonumber \\
W_{\pm} & = & 
\frac{\Delta^2}{3}\left( \frac{m}{k_{\pm}} \right)^2 q^2,
\nonumber \\
V_{\pm} & = &  
\frac{q^2}{2} \left( \frac{m}{k_{\pm}}\right) 
\biggl[\frac{\omega^2}{\sqrt{\omega^2 + \Delta^2} }
\mp \omega \biggr].
\nonumber
\end{eqnarray}

In case of $q \ne 0$ four linear independent solutions 
of Eq.~(\ref{Schrod}) have the  form:

\noindent
i) solution $u_{+}(z)$:
\[
u_{+}(z) = \left( 
{ \ e^{ik_{+}z + i\eta_{+}^{(+)}(z)} } \atop 
{ \alpha e^{ik_{+}z + i\eta_{-}^{(+)}(z)} } \right),
\]
where
\begin{eqnarray}
\eta_{+}^{(+)}(z) & = &
-\ i \lambda_{+} W_{+} z^3 + i \lambda_{+} V_{+} z^2
\nonumber \\
& & + \ 2i \lambda_{+}^2 V_{+} z + 2i \lambda_{+}^3 V_{+}
\nonumber \\
& \equiv & \tau_3^+ z^3 + \tau_2^+ z^2 + \tau_1^+ z + \tau_0^+,
\nonumber \\
\eta_{-}^{(+)}(z) & = &
\frac{1}{\alpha^2}\ \eta_{+}^{(+)}(z) + 
\frac{\Delta}{3\alpha}\left( \frac{m}{k_+} \right) q^2 z^3
\nonumber \\
& \equiv & \tau_3^- z^3 + \tau_2^- z^2 + \tau_1^- z + \tau_0^-;
\nonumber
\end{eqnarray}

\noindent
ii) solution $u_{-}(z)$:
\[
u_{-}(z) = \left( 
{\ e^{-ik_{+}z - i\eta_{+}^{(-)}(z)} } \atop 
  \alpha {e^{-ik_{+}z - i\eta_{-}^{(-)}(z)} } \right),
\]
where
\begin{eqnarray}
\eta_{\pm}^{(-)}(z) & = &
- \eta_{\pm}^{(+)}(-z)  
\nonumber \\
& = & \tau^{\pm}_3 z^3 - \tau_2^{\pm} z^2 + \tau_1^{\pm} z - \tau_0^{\pm};
\nonumber 
\end{eqnarray}

\noindent
iii) solution $v_{+}(z)$:
\[
v_{+}(z) = \left( 
{ \alpha e^{ik_{-}z + i\zeta_{+}^{(+)}(z)} } \atop 
{ \ e^{ik_{-}z + i\zeta_{-}^{(+)}(z)} } \right),
\]
where
\begin{eqnarray}
\zeta_{+}^{(+)}(z) & = &
\ i \lambda_{-} W_{-} z^3 + i \lambda_{-} V_{-} z^2
\nonumber \\
& & - \ 2i \lambda_{-}^2 V_{-} z + 2i \lambda_{-}^3 V_{-}
\nonumber \\
& \equiv & \rho_3^+ z^3 + \rho_2^+ z^2 + \rho_1^+ z + \rho_0^+,
\nonumber \\
\zeta_{-}^{(+)}(z) & = &
 \alpha^2 \zeta_{+}^{(+)}(z) + 
\frac{\alpha\Delta}{3}\left( \frac{m}{k_-} \right) q^2 z^3
\nonumber \\
& \equiv & \rho_3^- z^3 + \rho_2^- z^2 + \rho_1^- z + \rho_0^-;
\nonumber
\end{eqnarray}

\noindent
iv) solution $v_{-}(z)$:
\[
v_{-}(z) = \left( 
{\alpha e^{-ik_{-}z - i\zeta_{+}^{(-)}(z)} } \atop 
{\ e^{-ik_{-}z - i\zeta_{-}^{(-)}(z)} } \right),
\]
where
\begin{eqnarray}
\zeta_{\pm}^{(-)}(z) & = &
- \zeta_{\pm}^{(+)}(-z)  
\nonumber \\
& = & \rho^{\pm}_3 z^3 - \rho_2^{\pm} z^2 + \rho_1^{\pm} z - \rho_0^{\pm}.
\nonumber 
\end{eqnarray}

\end{document}